%% file: interpretA_w_.tex
\documentclass{jicspack}
\input{config}

\usepackage{enumerate}
\usepackage{graphicx}

\usepackage{amssymb}

\begin{document}

\begin{premaker}


\title{Case Study of  the Proof of \textit{Cook’s theorem}    \\ - Interpretation of $A(w)$ 
}
\author{Yu LI}
\ead{yu.li@u-picardie.fr}
\address{MIS, Universit\'{e} de Picardie Jules Verne, 33 rue Saint-Leu, 80090 Amiens, France}


\begin{abstract}

\textit{Cook’s theorem}  is commonly expressed such as any polynomial time-verifiable problem can be reduced to the \textit{SAT} problem. The proof of \textit{Cook’s theorem} consists in  constructing a propositional formula $A(w)$ to simulate a computation of \textit{TM}, and such  $A(w)$  is claimed to be \textit{CNF} to represent a polynomial time-verifiable problem $w$. In this paper, we investigate   $A(w)$ through a very simple example and show that, $A(w)$ has just  an appearance of \textit{CNF}, but not a true logical form.  This case study suggests that there exists the  \textit{begging the question} in \textit{Cook’s theorem}.

\end{abstract}
\end{premaker}


 \section{Introduction}
 
\textit{Cook’s theorem} \cite{cook1} is now expressed as any polynomial time-verifiable problem can be reduced to the  \textit{SAT (SATisfiability)}   problem. The proof of \textit{Cook’s theorem}  consists in simulating a computation of   \textit{TM (Turing Machine)}  by constructing a propositional formula $A(w)$ that is claimed to be  \textit{CNF (Conjonctive Normal Form)}   to represent the polynomial time-verifiable problem [1].
 
In this paper we investigate whether this $A(w)$ is a true logical form to represent  a problem through a very simple example.

 \section{Example }
 
  \subsection{ Polynomial time-verifiable problem and Turing Machine }
  
 A polynomial time-verifiable problem refers to a problem $w$ for which there exists a \textit{Turing Machine}  $M$  to verify a certificat $u$ in polynomial time, that is, check whether $u$ is a solution to $w$. 
 
Let us study a very simple polynomial time-verifiable problem :

Given a propositional formula   $w=\neg x$ for which there exists a \textit{Turing Machine}   $M$  to verify whether a truth value $u$ of $x$ is a solution to $w$. 

The  transition function  of  $M$  can be represented as follows:

\begin{tabular}{llllll}
   $q_0$ & 0 &  $\rightarrow$  &1& $N$& $q_1$ \\
    $q_0$ & 1 &  $\rightarrow$  &0& $N$& $q_1$ \\
    $q_1$ & 1 &  $\rightarrow$  &1& $R$& $q_Y$ \\
    $q_1$ & 0 &  $\rightarrow$  &0& $R$& $q_N$ \\
   \end{tabular}

where $N$ means that the tape head does not move, and $R$ means that the tape head  moves to right;  $q_Y$ refers to the state where $M$ stops and indicates that $u$ is a solution to $w$, and $q_N$ refers to the state where $M$ stops and indicates that $u$ is not a solution to $w$. 

  \subsection{Computation of Turing Machine }
  
A computation of  $M$ consists of a sequence of  configurations: $C(1), C(2), ..., C(T)$, where $T=Q(\mid w \mid)$ and $Q(n)$ is a polynomial function. A configuration $C(t)$  represents the situation of $M$ at time $t$ where  $M$ is in a state, with some symbols on its tape, with its head scanning a square, and the next configuration is determined by the transition function of $M$.

 Fig.1 and Fig. 2 illustrate two computations  of $M$ on inputs : $x=0$ and  $x=1$. 
  
    \begin{figure} [h]
\begin{center}
\includegraphics[scale=.3]{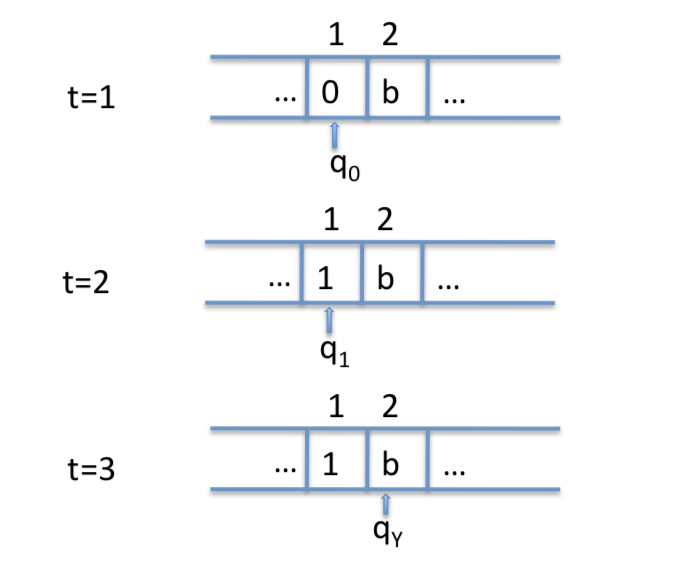}
\end{center}
\caption{The computation on input $x=0$.}
\label{fig1}
\end{figure}
  
  \begin{figure} [h]
\begin{center}
\includegraphics[scale=.3]{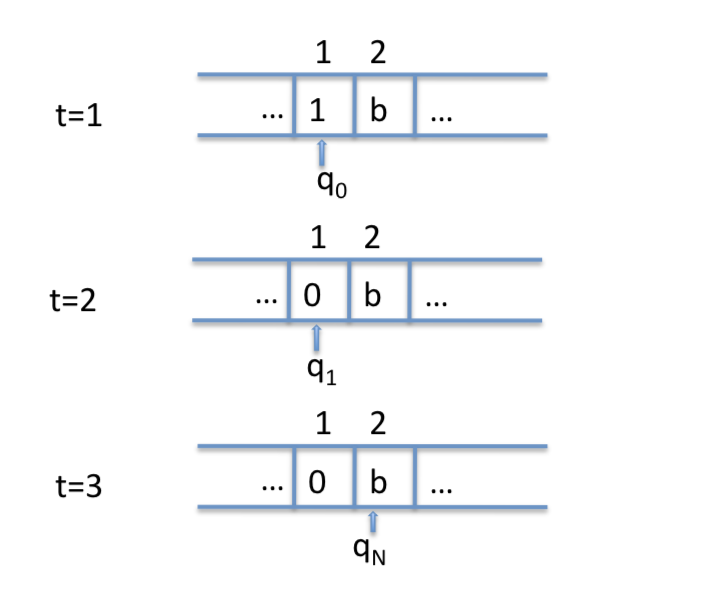}
\end{center}
\caption{The computation on input $x=1$.}
\label{fig2}
\end{figure}

\section{Form of $A(w)$ }

According to the proof of \textit{Cook’s theorem} \cite{cook1}\cite{garey}, the formula  $A(w)$  is built by simulating a computation of $M$, such as $A(w)=B \land  C  \land  D \land  E \land  F \land  G  \land  H \land  I $.  $A(w)$ is  claimed to represent a problem $w$. 

We construct $A(w)$ for the above example.
 
  \subsection{Basic elements}

   The machine $M$  possesses:
\begin{itemize}
 \item  4 states : $\{q_0, q_1, q_2=q_Y, q_3=q_N \}$, where $q_0$ is the initial state, and $q_2$,  $q_3$ are two final states.
 \item 3 symbols : $\{ \sigma_1=b, \sigma_2=0, \sigma_3=1 \}$, where  $\sigma_1$  is the blank symbol.
 \item  2 square numbers : $\{s=1, s=2\}$.
  \item 4 rules.
   \item $n$ is the  input size, $n=2$; $p(n)$ is a polynomial function of $n$, and $p(2)=3$.
 \item 3 times ($t=1, t=2, t=3$) et 2 steps to verify a certificat $u$ of $w$,   where $t=1$  corresponds to the time for the initial state of the machine.

 \end{itemize}

  \subsection{Proposition symbols}
  
  Three types of proposition symbols to represent a configuration of $M$ :

 \begin{itemize}
\item $P_{s,t}^i$  for $1\leq i \leq 3$, $1\leq s \leq 2$, $1\leq t \leq 3$. $P_{s,t}^i$  is true iff at step $t$  the  square number $s$ contains the symbol $\sigma_i$. 
\item  $Q_{t}^i$ for $1 \leq i \leq 4$, $1\leq t \leq 3 $. $Q_{t}^i$ is true iff at step t the machine is in state $q_i$.
\item  $S_{s,t}$ for $1\leq s \leq 2$, $1\leq t \leq 3$ is true iff at step $t$ the tape head scans square number $s$.
\end{itemize}

   \subsection{Propositions}

 1. $E={E_1   \land  E_2   \land   E_3}$, where $E_t$ represents the truth values of $P_{s,t}^i$, $Q_{t}^i$ and $S_{s,t}$ at time $t$: 
  \begin{itemize}
  \item $E_1 = Q_1^0   \land   S_{1,1}   \land   P_{1,1}^{2}  \land   P_{2,1}^{1}$  ($x =  0 (\sigma_{2}$)); $E_1 = Q_1^0   \land   S_{1,1}   \land   P_{1,1}^{3}  \land   P_{2,1}^{1}$ ($x =  1 (\sigma_{3}$))
   \item $E_2$ and  $E_2$ are determined by the transition function of $M$
   \end{itemize}

  2. $B={B_1   \land  B_2   \land   B_3}$, where $B_t$ asserts that at time $t$ one and only one square is scanned :

 \begin{itemize}
\item $B_1 = (S_{1,1}  \lor  S_{2,1})    \land ( \neg S_{1,1} \lor  \neg S_{2,1})$
\item $B_2 = (S_{1,2}  \lor  S_{2,2})    \land ( \neg S_{1,2} \lor  \neg S_{2,2})$
 \item$B_3 = (S_{1,3}  \lor  S_{2,3})    \land ( \neg S_{1,3} \lor  \neg S_{2,3})$
 \end{itemize}
  

3. $C={C_1   \land  C_2   \land   C_3}$,  where $C_t$ asserts  that at time $t$   there is one and only one symbol at each square. $C_t$ is the conjunction of all the $C_{i,t}$.

$C_1 = C_{1,1}  \land   C_{2,1} $:
 \begin{itemize}
\item $C_{1,1} = ( P_{1,1}^1  \lor  P_{1,1}^2 \lor P_{1,1}^3)    \land ( \neg P_{1,1}^1 \lor  \neg P_{1,1}^2)  \land ( \neg P_{1,1}^1 \lor  \neg P_{1,1}^3) \land ( \neg P_{1,1}^2 \lor  \neg P_{1,1}^3)$
\item $C_{2,1} = ( P_{2,1}^1  \lor  P_{2,1}^2 \lor P_{2,1}^3)    \land ( \neg P_{2,1}^1 \lor  \neg P_{2,1}^2)  \land ( \neg P_{2,1}^1 \lor  \neg P_{2,1}^3) \land ( \neg P_{2,1}^2 \lor  \neg P_{2,1}^3)$
 \end{itemize}

 $C_2 = C_{1,2}  \land   C_{2,2} $:
 \begin{itemize}
\item $C_{1,2} = ( P_{1,2}^1  \lor  P_{1,2}^2 \lor P_{1,2}^3)    \land ( \neg P_{1,2}^1 \lor  \neg P_{1,2}^2)  \land ( \neg P_{1,2}^1 \lor  \neg P_{1,2}^3) \land ( \neg P_{1,2}^2 \lor  \neg P_{1,2}^3)$
\item $C_{2,2} = ( P_{2,2}^1  \lor  P_{2,2}^2 \lor P_{2,2}^3)    \land ( \neg P_{2,2}^1 \lor  \neg P_{2,2}^2)  \land ( \neg P_{2,2}^1 \lor  \neg P_{2,2}^3) \land ( \neg P_{2,2}^2 \lor  \neg P_{2,2}^3)$
 \end{itemize}
 
  $C_3 = C_{1,3}  \land   C_{2,3} $:
 \begin{itemize}
\item $C_{1,3} = ( P_{1,3}^1  \lor  P_{1,3}^2 \lor P_{1,3}^3)    \land ( \neg P_{1,3}^1 \lor  \neg P_{1,3}^2)  \land ( \neg P_{1,3}^1 \lor  \neg P_{1,3}^3) \land ( \neg P_{1,3}^2 \lor  \neg P_{1,3}^3)$
\item $C_{2,3} = ( P_{2,3}^1  \lor  P_{2,3}^2 \lor P_{2,3}^3)    \land ( \neg P_{2,3}^1 \lor  \neg P_{2,3}^2)  \land ( \neg P_{2,3}^1 \lor  \neg P_{2,3}^3) \land ( \neg P_{2,3}^2 \lor  \neg P_{2,3}^3)$
 \end{itemize}


 4. $D={D_1  \land  D_2   \land   D_3}$,  where $D_t$ asserts that at time $t$ the machine is in one and only one state.

 \begin{itemize}
\item $D_1 = (Q_{1}^0  \lor  Q_{1}^1 \lor  Q_{1}^2 \lor  Q_{1}^3)    \land ( \neg Q_{1}^0 \lor  \neg Q_{1}^1)  \land ( \neg Q_{1}^0 \lor  \neg Q_{1}^2)  \land ( \neg Q_{1}^0 \lor  \neg Q_{1}^3)  \land ( \neg Q_{1}^1 \lor  \neg Q_{1}^2)  \land ( \neg Q_{1}^1 \lor  \neg Q_{1}^3)  \land ( \neg Q_{1}^2 \lor  \neg Q_{1}^3)$

\item $D_2 = (Q_{2}^0  \lor  Q_{2}^1 \lor  Q_{2}^2 \lor  Q_{2}^3)    \land ( \neg Q_{2}^0 \lor  \neg Q_{2}^1)  \land ( \neg Q_{2}^0 \lor  \neg Q_{2}^2)  \land ( \neg Q_{2}^0 \lor  \neg Q_{2}^3)  \land ( \neg Q_{2}^1 \lor  \neg Q_{2}^2)  \land ( \neg Q_{2}^1 \lor  \neg Q_{2}^3)  \land ( \neg Q_{2}^2 \lor  \neg Q_{2}^3)$

\item $D_3 = (Q_{3}^0  \lor  Q_{3}^1 \lor  Q_{3}^2 \lor  Q_{3}^3)    \land ( \neg Q_{3}^0 \lor  \neg Q_{3}^1)  \land ( \neg Q_{3}^0 \lor  \neg Q_{3}^2)  \land ( \neg Q_{3}^0 \lor  \neg Q_{3}^3)  \land ( \neg Q_{3}^1 \lor  \neg Q_{3}^2)  \land ( \neg Q_{3}^1 \lor  \neg Q_{3}^3)  \land ( \neg Q_{3}^2 \lor  \neg Q_{3}^3)$

 \end{itemize}
 


 5. $F$, $G$, and $H$ assert that for each time $t$ the values of the $P_{s,t}^i$, $Q_{t}^i$ and $S_{s,t}$ are updated properly. 
  
 $F={F_1   \land   F_2}$, where $F_t$ is the conjunction over all $i$ and $j$ of $F_{i,j}^t$, where $F_{i,j}^t$ asserts that at time $t$ the machine is in state $q_i$ scanning symbol $\sigma_j$, then at time $t+1$ $\sigma_j$ is changed into $\sigma_l$, where $\sigma_l$ is the symbol given by the transition function for $M$.
 
 $F_1=F_{0,2}^1   \land  F_{0,3}^1$ :
 \begin{itemize}
\item $F_{0,2}^1 =  ( \neg Q_1^0  \lor  \neg S_{1,1}   \lor  \neg P_{1,1}^2   \lor P_{1,2}^3)$, with the rule $(q_0, 0 \rightarrow 1, N, q_1)$ 
\item $F_{0,3}^1 =  ( \neg Q_1^0  \lor  \neg S_{1,1}   \lor  \neg P_{1,1}^3   \lor P_{1,2}^2)$, with the rule $(q_0, 1 \rightarrow 0, N, q_1)$ 
  \end{itemize}
  
  $F_2=F_{1,2}^2   \land  F_{1,3}^2$ :
 \begin{itemize}
\item $F_{1,2}^2 =  ( \neg Q_2^1  \lor  \neg S_{1,2}   \lor  \neg P_{1,2}^2   \lor P_{1,2}^2)$, with the rule $(q_1, 0 \rightarrow 0, R, q_N)$ 
\item $F_{1,3}^2 =  ( \neg Q_2^1  \lor  \neg S_{1,2}   \lor  \neg P_{1,2}^3   \lor P_{1,3}^3)$, with the rule $(q_1, 1 \rightarrow 1, R, q_Y)$ 
  \end{itemize}

 $G={G_1   \land   G_2}$, where $G_t$ is the conjunction over all $i$ and $j$ of $G_{i,j}^t$, where $G_{i,j}^t$ asserts that at time $t$ the machine is in state $q_i$ scanning symbol $\sigma_j$, then at time $t+1$ the machine is in state $q_k$, where $q_k$ is the state given by the transition function for $M$.
 
 $G_1=G_{0,2}^1   \land  G_{0,3}^1$ :
 \begin{itemize}
\item $G_{0,2}^1 =  ( \neg Q_1^0  \lor  \neg S_{1,1}   \lor  \neg P_{1,1}^2   \lor Q_{2}^1)$, with the rule $(q_0, 0 \rightarrow 1, N, q_1)$ 
\item $G_{0,3}^1 =  ( \neg Q_1^0  \lor  \neg S_{1,1}   \lor  \neg P_{1,1}^3   \lor Q_{2}^1)$, with the rule $(q_0, 1 \rightarrow 0, N, q_1)$ 
  \end{itemize}
  
  $G_2=G_{1,2}^2   \land  G_{1,3}^2$ :
 \begin{itemize}
\item $G_{1,2}^2 =  ( \neg Q_2^1  \lor  \neg S_{1,2}   \lor  \neg P_{1,2}^2   \lor Q_{3}^3)$, with the rule $(q_1, 0 \rightarrow 0, R, q_N)$ 
\item $G_{1,3}^2 =  ( \neg Q_2^1  \lor  \neg S_{1,2}   \lor  \neg P_{1,3}^3   \lor Q_{3}^2)$, with the rule $(q_1, 1 \rightarrow 1, R, q_Y)$ 
  \end{itemize}

$H={H_1   \land   H_2}$, where $H_t$ is the conjunction over all $i$ and $j$ of $G_{i,j}^t$, where $H_{i,j}^t$ asserts that at time $t$ the machine is in state $q_i$ scanning symbol $\sigma_j$, then at time $t+1$  the tape head moves  according to the transition function for $M$.
 
 $H_1=H_{0,2}^1   \land  H_{0,3}^1$ :
 \begin{itemize}
\item $H_{0,2}^1 =  ( \neg Q_1^0  \lor  \neg S_{1,1}   \lor  \neg P_{1,1}^2   \lor S_{1,2})$, with the rule $(q_0, 0 \rightarrow 1, N, q_1)$ 
\item $H_{0,3}^1 =  ( \neg Q_1^0  \lor  \neg S_{1,1}   \lor  \neg P_{1,1}^3   \lor S_{1,2})$, with the rule $(q_0, 1 \rightarrow 0, N, q_1)$ 
  \end{itemize}
  
  $H_2=H_{1,2}^2   \land  H_{1,3}^2$ :
 \begin{itemize}
\item $H_{1,2}^2 =  ( \neg Q_2^1  \lor  \neg S_{1,2}   \lor  \neg P_{1,2}^2   \lor S_{2,3})$, with the rule $(q_1, 0 \rightarrow 0, R, q_N)$ 
\item $H_{1,3}^2 =  ( \neg Q_2^1  \lor  \neg S_{1,2}   \lor  \neg P_{1,3}^3   \lor S_{2,3})$, with the rule $(q_1, 1 \rightarrow 1, R, q_Y)$ 
  \end{itemize}


6.  $I = (Q_{3}^2  \lor    Q_{3}^3) \land  (Q_{3}^2   \lor  \neg   Q_{3}^3) \land  (\neg Q_{3}^2   \lor  Q_{3}^3)$, asserts that the machine reaches the state $q_y$ or $q_N$  at time 3. \\
  
Finally, $A(w)=B \land  C  \land  D \land  E \land  F \land  G  \land  H \land  I $.
 
 \section{ Conjunctive form of $A(w)$} 
 
We develop $A(w)$  as a computation of $M$ for  $x=0$ as input    (see Fig. 1) in order to clarify  the real sense of  $A(w)$.

 Let us define the configuration and  the transition of configurations of $M$ :

$C(t)$ :  the truth values of  $P_{s,t}^i$, $Q_{t}^i$, $S_{s,t}$ and their constraints. 

$C(t)   \rightarrow C(t+1)$ :  $C(t)$ is changed to $C(t+1)$  according to the transition function of $M$.  \\

1. At $t=1$, $C(1) = E_1  \land B_1  \land C_1  \land D_1$ :

   \begin{figure} [h]
\begin{center}
\includegraphics[scale=.7]{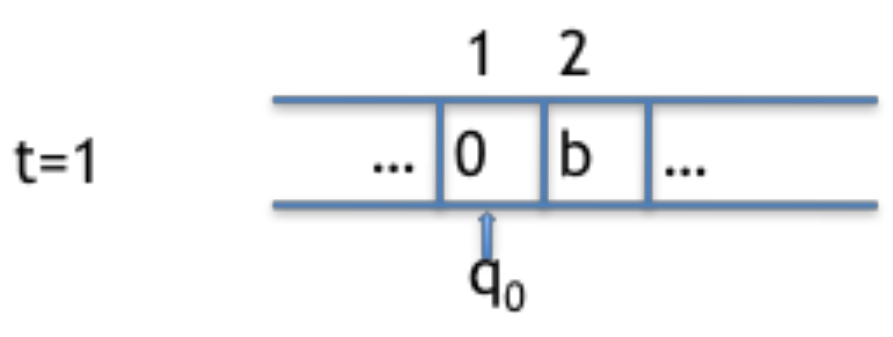}
\end{center}
\label{fig1.1}
\end{figure}

\begin{itemize}
\item $E_1 = Q_1^0   \land   S_{1,1}   \land   P_{1,1}^{2}  \land   P_{2,1}^{1}$, representing the initial configuration where  $M$ is in  $q_0$,  the tape head scans the square of number 1, and a string $0b$ is on the tape.

\item $B_1 = (S_{1,1}  \lor  S_{2,1})    \land ( \neg S_{1,1} \lor  \neg S_{2,1}) $.

\item $C_1 = C_{1,1}  \land   C_{2,1} $:

 \begin{itemize}
\item $C_{1,1} = ( P_{1,1}^1  \lor  P_{1,1}^2 \lor P_{1,1}^3)    \land ( \neg P_{1,1}^1 \lor  \neg P_{1,1}^2)  \land ( \neg P_{1,1}^1 \lor  \neg P_{1,1}^3) \land ( \neg P_{1,1}^2 \lor  \neg P_{1,1}^3) $
\item $C_{2,1} = ( P_{2,1}^1  \lor  P_{2,1}^2 \lor P_{2,1}^3)    \land ( \neg P_{2,1}^1 \lor  \neg P_{2,1}^2)  \land ( \neg P_{2,1}^1 \lor  \neg P_{2,1}^3) \land ( \neg P_{2,1}^2 \lor  \neg P_{2,1}^3) $
 \end{itemize}
 
\item $D_1 = (Q_{1}^0  \lor  Q_{1}^1 \lor  Q_{1}^2 \lor  Q_{1}^3)    \land ( \neg Q_{1}^0 \lor  \neg Q_{1}^1)  \land ( \neg Q_{1}^0 \lor  \neg Q_{1}^2)  \land ( \neg Q_{1}^0 \lor  \neg Q_{1}^3)  \land ( \neg Q_{1}^1 \lor  \neg Q_{1}^2)  \land ( \neg Q_{1}^1 \lor  \neg Q_{1}^3)  \land ( \neg Q_{1}^2 \lor  \neg Q_{1}^3)$
 \end{itemize}

2.  At $t=2$,  $C(2) = E_2  \land B_2  \land C_2  \land D_2$ is obtained from $C(1)  \land (C(1)   \rightarrow C(2))$.
 
    \begin{figure} [h]
\begin{center}
\includegraphics[scale=.3]{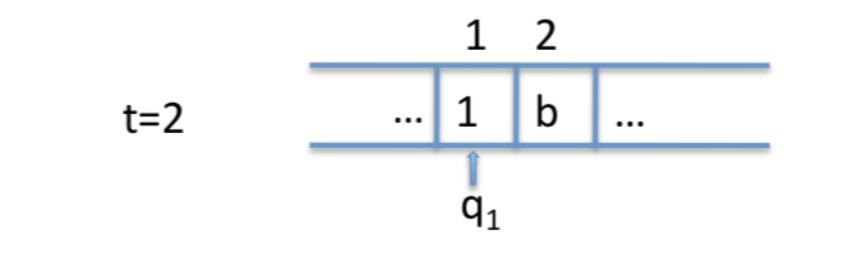}
\end{center}
\label{fig1.1}
\end{figure}

$C(1) \rightarrow C(2)$ is represented by $F$, $G$ and $H$ at $t=1$ :
 
  - $F_{0,2}^1 =  ( \neg Q_1^0  \lor  \neg S_{1,1}   \lor  \neg P_{1,1}^2   \lor P_{1,2}^3)$, with the rule $(q_0, 0 \rightarrow 1, N, q_1)$ 
 
 - $G_{0,2}^1 =  ( \neg Q_1^0  \lor  \neg S_{1,1}   \lor  \neg P_{1,1}^2   \lor Q_{2}^1)$, with the rule $(q_0, 0 \rightarrow 1, N, q_1)$ 
  
-  $H_{0,2}^1 =  ( \neg Q_1^0  \lor  \neg S_{1,1}   \lor  \neg P_{1,1}^2   \lor S_{1,2})$, with the rule $(q_0, 0 \rightarrow 1, N, q_1)$ 
  
 \begin{itemize}
\item $E_2 = Q_2^1  \land   S_{1,1}   \land  P_{1,2}^{3}  \land   P_{2,2}^{1}$,  with  $Q_2^1=1$,     $S_{1,2}=1$,     $P_{1,2}^{3}=1$,     $P_{2,2}^{1}=1$, and other proposition symbols concerning $t=2$ are assigned with 0.

\item $B_2 = (S_{1,2}  \lor  S_{2,2})    \land ( \neg S_{1,2} \lor  \neg S_{2,2})$

\item $C_2 = C_{1,2}  \land   C_{2,2} $:
 \begin{itemize}
\item $C_{1,2} = ( P_{1,2}^1  \lor  P_{1,2}^2 \lor P_{1,2}^3)    \land ( \neg P_{1,2}^1 \lor  \neg P_{1,2}^2)  \land ( \neg P_{1,2}^1 \lor  \neg P_{1,2}^3) \land ( \neg P_{1,2}^2 \lor  \neg P_{1,2}^3) $
\item $C_{2,2} = ( P_{2,2}^1  \lor  P_{2,2}^2 \lor P_{2,2}^3)    \land ( \neg P_{2,2}^1 \lor  \neg P_{2,2}^2)  \land ( \neg P_{2,2}^1 \lor  \neg P_{2,2}^3) \land ( \neg P_{2,2}^2 \lor  \neg P_{2,2}^3) $
 \end{itemize}
 
\item  $D_2 = (Q_{2}^0  \lor  Q_{2}^1 \lor  Q_{2}^2 \lor  Q_{2}^3)    \land ( \neg Q_{2}^0 \lor  \neg Q_{2}^1)  \land ( \neg Q_{2}^0 \lor  \neg Q_{2}^2)  \land ( \neg Q_{2}^0 \lor  \neg Q_{2}^3)  \land ( \neg Q_{2}^1 \lor  \neg Q_{2}^2)  \land ( \neg Q_{2}^1 \lor  \neg Q_{2}^3)  \land ( \neg Q_{2}^2 \lor  \neg Q_{2}^3)$

 \end{itemize} 

3.  At $t=3$,  $C(3) = E_3  \land B_3  \land C_3 \land D_3$  is obtained from  $C(2)  \land (C(2)   \rightarrow C(3))$.
  
    \begin{figure} [h]
\begin{center}
\includegraphics[scale=.3]{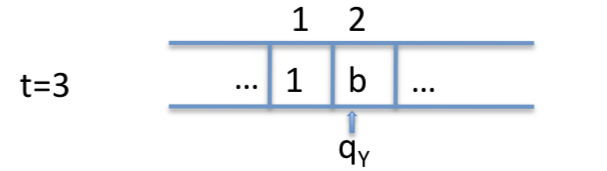}
\end{center}
\label{fig1.1}
\end{figure}

$C(2) \rightarrow C(3)$ is represented by $F$, $G$ and $H$ at $t=2$ :

 $F_{1,3}^2 =  ( \neg Q_2^1  \lor  \neg S_{1,2}   \lor  \neg P_{1,2}^3   \lor P_{1,3}^3)$, with the rule $(q_1, 1 \rightarrow 1, R, q_Y)$ 
 
 $G_{1,3}^2 =  ( \neg Q_2^1  \lor  \neg S_{1,2}   \lor  \neg P_{1,3}^3   \lor Q_{3}^2)$, with the rule $(q_1, 1 \rightarrow 1, R, q_Y)$ 
 
 $H_{1,3}^2 =  ( \neg Q_2^1  \lor  \neg S_{1,2}   \lor  \neg P_{1,3}^3   \lor S_{2,3})$, with the rule $(q_1, 1 \rightarrow 1, R, q_Y)$ 
 
 \begin{itemize}

\item $E_3 = Q_3^2  \land  S_{2,3}   \land  P_{1,3}^{3}  \land   P_{2,3}^{1}$,  with $Q_3^2=1$,     $S_{2,3}=1$,     $P_{1,3}^{3}=1$,     $P_{2,3}^{1}=1$ , and other proposition symbols concerning $t=3$ are assigned with 0.

\item $B_3 = (S_{1,3}  \lor  S_{2,3})    \land ( \neg S_{1,3} \lor  \neg S_{2,3})$

\item  $C_3 = C_{1,3}  \land   C_{2,3} $:
 \begin{itemize}
\item $C_{1,3} = ( P_{1,3}^1  \lor  P_{1,3}^2 \lor P_{1,3}^3)    \land ( \neg P_{1,3}^1 \lor  \neg P_{1,3}^2)  \land ( \neg P_{1,3}^1 \lor  \neg P_{1,3}^3) \land ( \neg P_{1,3}^2 \lor  \neg P_{1,3}^3)$
\item $C_{2,3} = ( P_{2,3}^1  \lor  P_{2,3}^2 \lor P_{2,3}^3)    \land ( \neg P_{2,3}^1 \lor  \neg P_{2,3}^2)  \land ( \neg P_{2,3}^1 \lor  \neg P_{2,3}^3) \land ( \neg P_{2,3}^2 \lor  \neg P_{2,3}^3)$
 \end{itemize}
 
\item  $D_3 = (Q_{3}^0  \lor  Q_{3}^1 \lor  Q_{3}^2 \lor  Q_{3}^3)    \land ( \neg Q_{3}^0 \lor  \neg Q_{3}^1)  \land ( \neg Q_{3}^0 \lor  \neg Q_{3}^2)  \land ( \neg Q_{3}^0 \lor  \neg Q_{3}^3)  \land ( \neg Q_{3}^1 \lor  \neg Q_{3}^2)  \land ( \neg Q_{3}^1 \lor  \neg Q_{3}^3)  \land ( \neg Q_{3}^2 \lor  \neg Q_{3}^3)$ \\

  \end{itemize}

Therefore, the computation of $M$ for $x=0$ as input can be represented as :

$
C(1)  \land  (C(1) \rightarrow C(2))  \land (C(2) \rightarrow C(3))   \\
=  C(1)  \land C(2) \land C(3)  \\
= (E_1  \land B_1  \land C_1  \land D_1)  \land   (E_2  \land B_2  \land C_2  \land D_2) \land (E_3  \land B_3  \land C_3 \land D_3)\\
=    E  \land B  \land C  \land D  \\
=  A(w)
$

It can be seen that $A(w)$ is just the conjonction of all configurations of $M$ to simulate a concret computation of $M$ for verifying a certificat $u$ of $w$. Given an input $u$ ($x=0$ or $x=1$ in this example), whether $M$ accepts it or not,  $A(w)$ is always true. Obviously, $A(w)$ has just an appearance of conjunctive form, but not a true logical form.

  \section{Conclusion} 
  
   In fact, a true  \textit{CNF} formula   is implied in the transition function of $M$ corresponding to $F$, $G$, $H$ as well as $ C(t) \rightarrow C(t+1)$, however the transition function  of $M$ is based on the expressible logical structure of a problem.

 Therefore, it is not that any polynomial time-verifiable problem can be reduced to the SAT problem, but any polynomial time-verifiable problem itself asserts that such problem is representable by a CNF formula. In other words,  there exists the \textit{begging the question} in \textit{Cook’s theorem}. 
   
   \section*{Acknowledgements}
Thanks to Mr Chumin LI for his suggestion to use this simple example to study $A(w)$.

\end{document}

%% file: config.tex
\setvolume{8}                             
\setyear{2012}                             
\setpagerange{1}{8}                    
\setheadauthor{X. Liu et al.}          
\setissn{1553--9105} \setpubdate{January 2012} \setno{1}

\afterpage{\beginheader}                   